\documentclass[aps,prb,twocolumn,groupedaddress]{revtex4}
\usepackage{epsfig}
\usepackage[american]{babel}
\usepackage{amsmath}
\usepackage[dvips]{color}
\usepackage[latin2]{inputenc}
\usepackage[T1]{fontenc}
\usepackage[american]{babel}
\usepackage{color}
\usepackage{graphicx}

\newcommand{\sech}{\operatorname{sech}}

\newcommand{\LG}{\mathbf {G}}

\begin{document}




\title{Diagrammatic content of the DMFT for the Holstein polaron problem in finite dimensions}

\author{O. S. Bari\v si\' c}


\affiliation{Institute of Physics, Bijeni\v cka c. 46, HR-10000 Zagreb, Croatia}

\begin{abstract}

In the context of the Holstein polaron problem it is shown that the dynamical mean field theory (DMFT) corresponds to the summation of a special class of local diagrams in the skeleton expansion of the self-energy. In the real space representation, these local diagrams are characterized by the absence of vertex corrections involving phonons at different lattice sites. Such corrections vanish in the limit of infinite dimensions, for which the DMFT provides the exact solution of the Holstein polaron problem. However, for finite dimensional systems the accuracy of the DMFT is limited. In particular, it cannot describe correctly the adiabatic spreading of the polaron over multiple lattice sites. Arguments are given that the DMFT limitations on vertex corrections found for the Holstein polaron problem persist for finite electron densities and arbitrary phonon dispersion.

\end{abstract}

\maketitle

Since the early works in which some important aspects of the Mott localization were accounted for successfully, the DMFT has been used for a broad range of strongly-correlated problems.\cite{Pruschke,Koller} In the DMFT the quasi-particle properties are calculated by treating the self-energy as a local ($k$ independent) quantity. This approach is motivated by the observation that for some models in the infinite dimensional limit $D\rightarrow\infty$ the exact self-energy is local. In particular, for the Hubbard model this property is derived from the diagrammatic perturbation theory in the interaction strength $U$,\cite{Metzner} as well as from the diagrammatic expansion around the atomic limit.\cite{Metzner2} Considerable attention was given to the electron-phonon Holstein model too,\cite{Paci} for which the nature of the perturbation theory is local for $D\rightarrow\infty$. There are however important differences between the two models, which are best illustrated by the fact that motivates the current study. Unlike in the Hubbard model, which leaves the first electron in the system free, renormalization occurs in the Holstein model irrespectively of the dimension $D$, provided that the phonon frequency is finite.

It is commonly believed that the DMFT results provide valuable insights on real materials, although for finite dimensional systems the level of approximation is frequently difficult to estimate. For these reasons it is particularly interesting to analyze the limitations of the DMFT in the context of the Holstein polaron problem because accurate results for low frequencies, at which the polaronic correlations are the strongest, are available in low dimensions for the whole parameter space.\cite{Filippis,Barisic5} Along these lines, in this work the diagrammatic content of the DMFT is analyzed first, in order to identify exactly which are the contributions ignored for finite $D$. In the next step, the physical meaning of these is investigated.

The Holstein model\cite{Holstein} describes the tight-binding electrons in the nearest-neighbor approximation, coupled to one branch of dispersionless optical phonons, 

\begin{eqnarray}
\hat H&=&-t\sum_{r,\delta}c_r^\dagger c_{r+\delta}+\omega_0\sum_rb_r^\dagger b_r \nonumber\\
&&-g\sum_r c_r^\dagger c_r(b_r^\dagger+ b_r)\;.
\label{Eq001}
\end{eqnarray}

\noindent Here, $c_r^\dagger$ and $b_r^\dagger$ are the creation operators for the electron and the phonon, respectively, $t$ is the electron hopping integral, $\omega_0$ is the phonon energy, and $g$ is the electron-phonon coupling constant. As the bare phonons and the coupling are local, the lattice geometry and dimensionality are expressed only through the first term in Eq.~(\ref{Eq001}) involving the summation over nearest-neighbor sites $\delta$. The spin index is omitted since only the single-electron (polaron) problem is considered here.

\begin{figure}[tb]

\begin{center}{\scalebox{1.1}
{\includegraphics{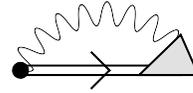}}}
\end{center}

\caption{The exact electron self-energy $\Sigma_k(\omega)$ for the Holstein polaron problem. The double line represent the exact electron propagator $G_k(\omega)$, the shaded triangle is the exact vertex and the single wavy line is the bare phonon propagator $D^{(0)}(\omega)$.\label{Fig001}}

\end{figure}

The exact single-electron propagator in the momentum representation is given by

\begin{equation}
G_k(\omega)=\frac{1}{\omega-\varepsilon_k-\Sigma_k(\omega)}\;,
\end{equation}

\noindent with $\varepsilon_k$ the free-electron energy, and $\Sigma_k(\omega)$ is the exact self-energy shown in the diagrammatic representation in Fig.~\ref{Fig001}. There are no contributions to the phonon propagator due to the creation of the electron-hole pairs, since $G_k(\omega)$ describes the dynamics when only one electron is intermittently added to the system.\cite{Ciuchi,BBarisic} Thus, unlike for finite density cases, the phonon line in Fig.~\ref{Fig001} represents the bare phonon propagator. In particular, for the Holstein model, this propagator is $k$-independent (local), $D^{(0)}(\omega)=2\omega_0/(\omega^2-\omega_0^2+i\eta\omega_0)$. The polaron effects on the phonon self-energy can be investigated by considering the case of one electron permanently present in the system, as discussed in Refs.~\onlinecite{BBarisic,Barisic3}.

For the Holstein polaron problem (\ref{Eq001}) the DMFT can be formulated as an iterative procedure of generating a hierarchy of diagrams in the perturbative expansion of the electron self-energy. In each iterative step, one first considers explicitly the electron-phonon interaction on one lattice site only, usually referred to as the impurity site. This permits the expression of the electron self-energy associated with the impurity site~$s$ in terms of local electron and phonon propagators. In particular, by using the bare phonon propagator to describe the phonons, the impurity self-energy for the Holstein polaron problem can be evaluated\cite{Ciuchi} in terms of the continued fraction

\begin{equation}
\Sigma^{(n)}(\omega)
=\frac{g^2}{[\mathcal{G}^{(n)}_{s,s}(\omega-\omega_0)]^{-1}-\frac{2g^2}
{[\mathcal{G}^{(n)} _{s,s} (\omega-2\omega_0)]^{-1}- ...}}\;,\label{SEImpurity}
\end{equation}

\noindent where $\mathcal{G}^{(n)}_{s,s}(\omega)$ is the impurity propagator corresponding to the $n$th iterative step. The translationally invariant DMFT electron propagator of the $n$th iteration is obtained by treating all lattice sites on equal footing using the self-energy $\Sigma^{(n)}(\omega)$ of Eq.(\ref{SEImpurity}) according to 

\begin{equation}
\LG_{i,j}^{(n)}(\omega)=G^0_{i,j}(\omega)+ \Sigma^{(n)}(\omega) 
\sum_rG^0_{i,r}(\omega)\LG_{r,j}^{(n)}(\omega)\;,\label{Gij0}
\end{equation}

\noindent with $G^0_{i,j}(\omega)$ the free electron propagator. By taking the Fourier transform of Eq.~(\ref{Gij0}), in the momentum space one obtains 

\begin{equation}
\left[\LG_k^{(n)}(\omega)\right]^{-1}=\left[G_k^0(\omega)\right]^{-1}-\Sigma^{(n)} (\omega)\;.\label{Gk0}
\end{equation}

In order to establish a direct connection with the diagrammatic theory, it is appropriate to start the DMFT iterative procedure with the local free-electron propagator $G^0_{s,s}(\omega)$ as the initial $n=1$ guess in Eq.~(\ref{SEImpurity}), i.e., $\mathcal{G}^{(1)}_{s,s}(\omega)=G^0_{s,s}(\omega)$. With this initial step, one obtains in Eq. (\ref{SEImpurity}) the self-energy $\Sigma^{(1)}(\omega)$ that characterizes the exact solution of the problem in which the electron couples with a single phonon mode at the impurity site,

\begin{equation}
G_{i,j}(\omega)=G^0_{i,j}(\omega)+\Sigma^{(1)}(\omega)\; 
G^0_{i,s}(\omega) \;G_{s,j}(\omega)\;.\label{MAGF}
\end{equation} 

\noindent By examining the diagrammatic expansion\cite{Ciuchi} of $\Sigma^{(1)}(\omega)$ in Eq.~(\ref{SEImpurity}) order by order in $g^2$, one can verify that $G_{i,j}(\omega)$ of Eq. (\ref{MAGF}) is the exact solution of the single-electron tunneling through a quantum dot involving a single phonon mode.

For $\Sigma^{(n)}(\omega)= \Sigma^{(1)}(\omega)$, the propagator (\ref{Gk0}) is equal\cite{Barisic6} to the electron propagator derived in the context of the momentum averaging (MA) approximation, investigated in Refs.~\onlinecite{Berciu,Goodvin}. It involves all the processes for which the phonons occupy just one lattice site at the same time. 

At the beginning of each $n>1$ iterative step, the new impurity propagator entering Eq. (\ref{SEImpurity}) is determined in terms of the self-energy considered in the previous step, 

\begin{equation}
\mathcal{G}^{(n)}_{s,s}(\omega)=G^0_{s,s}(\omega)+\Sigma^{(n-1)}(\omega) 
\sum_{r\neq s}G^0_{s,r}(\omega)\mathcal{G}_{r,s}^{(n)}(\omega)\;.\label{Gijnp1}
\end{equation}

\begin{figure}[tb]

\begin{center}{\scalebox{0.5}
{\includegraphics{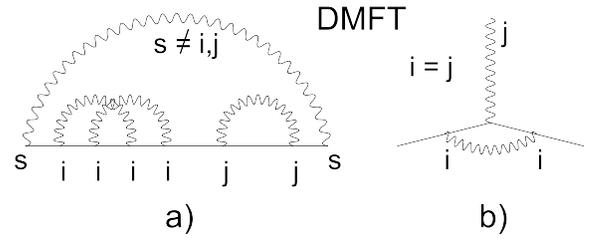}}}
\end{center}

\caption{In the DMFT context there is no crossing between phonon lines corresponding to different lattice sites in the diagrammatic expansion of the electron self-energy. The vertex corrections are limited to a single lattice site. \label{Fig002}}

\end{figure}

\noindent Such expression for $\mathcal{G}^{(n)}_{s,s}(\omega)$, which  is not translationally invariant by construction ($\Sigma^{(n)}(\omega)\neq0$), prevents the double counting of the diagrams.\cite{Jarrell} In particular, the self-energy diagram shown in Fig.~\ref{Fig002}a is generated in the second step ($n=2$) of the DMFT iterative procedure. The phonon lines corresponding to the sites $i,j\neq s$ describe the processes considered in the first step. These processes are included by $\mathcal{G}^{(2)}_{s,s}(\omega)$ in Eq. (\ref{Gijnp1}). The phonon line corresponding to the impurity site $s$ is added trough Eq. (\ref{SEImpurity}) by inserting $\mathcal{G}^{(2)}_{s,s}(\omega)$ into it.

One sees that each step of the DMFT iteration adds the phonon lines associated with the impurity site $s$ to the self-energy diagrams. In general, the diagrams obtained in the $n$th iterative step may involve phonons at up to $n$ different lattice sites at the same instant of time. However, because of the particular way in which the diagrams are generated, the phonon lines corresponding to different lattice sites never cross, i.e., the only vertex corrections considered by the DMFT involve one lattice site, as illustrated in Fig.~\ref{Fig002}b.

In actual calculations, the continued fraction (\ref{SEImpurity}) is evaluated to some finite order $M$, where $M$ defines the maximal number of phonon excitations associated with the impurity site.\cite{Ciuchi} For $M$ sufficiently large, $M\gtrsim g^2/\omega_0^2$, the higher order contributions in Eq.~(\ref{SEImpurity}) can be neglected, and one repeats the iterative procedure until the results converge to values satisfying a predetermined criteria. If $N$ is the number of iterative steps, the diagrams taken into account involve up to $M\times N$ phonon lines at the same instant of time.

For the initial impurity propagator $\mathcal{G}^{(1)}_{s,s}(\omega)$ in Eq.~(\ref{SEImpurity}), the local free-electron propagator $G^0_{s,s}(\omega)$ is used in order to identify exactly the diagrams contributing to the DMFT electron propagator. However, one can consider all other choices, assuming as usual in the DMFT that they converge to the same result. In fact, the DMFT is commonly iterated until the self-consistent solution,

\begin{equation}
\left[\mathcal{G}^{(n)}_{s,s}(\omega)\right]^{-1}=[\LG^{(n)}_{s,s}(\omega)]^{-1}+\Sigma^{(n)}(\omega)\;,
\end{equation} 

\noindent is achieved with no particular restrictions on the initial impurity propagator $\mathcal{G}^{(1)}_{s,s}(\omega)$. The relation between the DMFT and perturbation expansion established here shows that the existence of (at least one) self-consistent solution relies on the applicability of the perturbation series associated with the DMFT.

The absence of vertex corrections involving multiple lattice sites in the diagrammatic representation of the DMFT is not a limitation uniquely related to the single-electron problem. It persists for finite densities as well. Namely, in each DMFT iteration, one first calculates the impurity self-energy in terms of local propagators, taking into account the vertex corrections involving the impurity site only. This restriction on vertex corrections is not removed by restoring the translational symmetry in Eq.~(\ref{Gk0}), whatever the electron density. Furthermore, the above argument, based on the topology of the diagrammatic expansion, applies for dispersive phonons as well. 

For the Holstein polaron problem, the physical meaning of the diagrams neglected by the DMFT can be easily determined by comparing to the previous analytical and numerical results. First, it should be noticed that in the atomic (small $t$) limit, the exact electron self-energy $\Sigma_k(\omega)$ becomes $k$-independent (local) irrespectively of the dimensionality of the system. This is consistent with the DMFT. However, for finite dimensional cases, non-local contributions to the electron self-energy appear by increasing $t$. In particular, the limitations of the DMFT are observed best for large adiabatic polarons, for which the local and non-local contributions to the electron self-energy are equally important.\cite{BBarisic} For the Holstein model these polarons form in 1D for $t\gg\omega_0$ and $(t/\omega_0)^\frac{1}{4}\lesssim g/\omega_0\lesssim (t/\omega_0)^\frac{1}{2}$.\cite{BBarisic}

The original results for large adiabatic polarons were derived in pioneering works by applying the continuum adiabatic approximation,\cite{Rashba, Holstein} which breaks the translational symmetry from the outset. In this approximation, the electron wave function $\eta_r$ and the lattice deformation $u_r$ are obtained as

\begin{equation}
 u_r =\frac{2g}{\omega_0}\;|\eta_r |^2\;,\;\;\;\eta_r =
\frac{\sqrt\lambda}{2}\sech\left[\lambda\;(r-\xi/a)/2\right]\;,\label{HolLPol}
\end{equation}

\noindent where $u_r$ is the classical lattice deformation at the site $r$ in units of the space uncertainty of the zero-point motion,
$\lambda=g^2/t\;\omega_0$ defines the polaron size $d_{ad}\sim1/\lambda$ ($d_{ad}\gtrsim 1$), and $\xi/a$ is the position of the polaron along the continuum, with $a$ the lattice constant. The effective mass $m_{pol}$ of the polaron (\ref{HolLPol}) is given by the power law\cite{Holstein}

\begin{equation}
m_{pol}\sim\sum_r \left(\partial u_r/\partial\xi\right)^2\Rightarrow\;m_{pol}/m_{el}\sim\left(g/\omega_0\right)^4\lambda^2\;.\label{HolmPol}
\end{equation}

\noindent In the regime of large adiabatic polarons, Eq. (\ref{HolmPol}) reproduces well the exact effective mass derived either directly by the full diagrammatic summation\cite{BBarisic} or from the polaron band structure calculated by the relevant coherent states method (RCSM),\cite{Barisic5} $m_{el}/m_{pol}=\partial E_k/\partial\varepsilon_k|_{k=0}$, with $E_k$ the lowest polaron band dispersion. In other words, the effective mass that follows from the exact electron self-energy $\Sigma_k(\omega)$,\cite{Mahan}
 
\begin{equation}
\frac{m_{el}}{m_{pol}}=\frac{1+\partial_{k^2}\Sigma(k,\omega)|_{k=0}}
{1-\partial_\omega\Sigma(k,\omega)|_{\omega=E_0}}\;,\label{Eq06}
\end{equation}

\noindent should behave according to Eq.~(\ref{HolmPol}) in the regime of large adiabatic polarons. The denominator in Eq.~(\ref{Eq06}) is the inverse of the quasiparticle weight $Z_0$. From Eq.~(\ref{HolLPol}) one can estimate

\begin{equation}
\ln Z_0\sim\;-\sum_ru_r^2\sim\;-(g/\omega_0)^2\;\lambda\;,
\end{equation}

\noindent i.e., for $\lambda$ constant, $Z_0$ exponentially decreases with the coupling, $u_r\sim g/\omega_0$. Thus, according to Eq. (\ref{HolmPol}), for large adiabatic polarons the nonlocal contributions that determine the numerator in Eq.~(\ref{Eq06}) are exponentially large, just as are the contributions in the denominator. Obviously, it is not possible to achieve this result within the DMFT because the self-energy is local.

Additional important insights on the applicability of the DMFT can be obtained from the polaron binding energy. It is instructive to start the analysis with the weak-coupling $ g/\omega_0<(t/\omega_0)^\frac{1}{4}$ and the small polaron $\lambda\gg1$ limit, for which the DMFT converges to the exact solution.

\begin{figure}[tb]

\begin{center}{\scalebox{0.3}
{\includegraphics{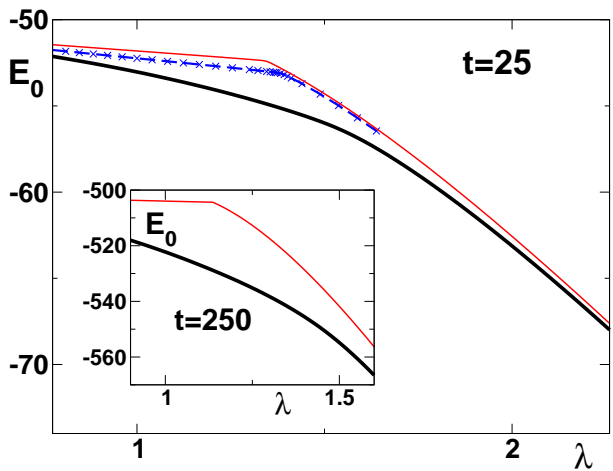}}}
\end{center}

\caption{(Color online) The RCSM (thick), DMFT (dashed with symbols), and MA (thin) curves are the polaron ground-state energies $E_0$ for $t=25$. The inset compares the MA and the RCSM for $t=250$. $\omega_0$ is used as the unit of energy.\label{Fig004}}

\end{figure}

Although Eq.~(\ref{HolLPol}) implies that the size of the adiabatic polaron in 1D increases infinitely as $\lambda$ decreases, it should be stressed that the upper limit on the range of adiabatic correlations is independent of the coupling constant $g$. It is given by the length $\sqrt{t/\omega_0}$ over which the free electron diffuses within a lattice period $1/\omega_0$.\cite{Emin} In particular, for $d_{ad}\sim1/\lambda\sim\sqrt{t/\omega_0}$, there is a smooth crossover between the large adiabatic and the nonadiabatic polarons, the latter corresponding to the weak-coupling limit.\cite{BBarisic} In the absence of adiabatic correlations for weak couplings, the exact electron self-energy is local and accurately reproduced by the DMFT. Accordingly, as the weak-coupling regime is approached by decreasing $\lambda$ [see the left side of Fig.~\ref{Fig004}], one observes that the deviations of the DMFT ground-state energy with respect to the RCSM results decrease. 

It can be argued that the main length scale over which the vertex corrections are important in the ground state energy $E_0$, is determined by adiabatic correlations. In particular, for low-frequencies $\omega\approx E_0$ this length scale is closely related to the size of the (adiabatic) polaron, scaling as $1/\lambda$. Indeed, on the right side of Fig.~\ref{Fig004} the DMFT approaches the RCSM ground-state energy as $\lambda$ is increased. However, the DMFT fails to describe the adiabatic spreading of small polarons and the adiabatic hopping to the neighboring sites. That is, on the right side of Fig.~\ref{Fig004} the MA, corresponding to the first step of the DMFT, exhibits inaccuracies similar to those of the DMFT. The DMFT is not a substantial improvement over the MA in the description of the adiabatic correlations since, in both cases, the vertex corrections involving more than one lattice site are neglected. It is worth noting in this respect that the MA approach can be improved by including all the lowest order vertex corrections.\cite{Berciu2} On the other hand, the DMFT is usually extended in the context of quantum cluster theories.\cite{Maier} In this latter case, the vertex corrections involving the sites of a chosen cluster, rather than a single site, are taken into consideration.

The central part of Fig.~\ref{Fig004} reveals that, in the regime between the weak-coupling and small-polaron limits, the DMFT and MA data show significant deviations from the RCSM ground-state energy. These deviations become more pronounced by increasing $t/\omega_0$, which can be seen from the comparison of the MA and the RCSM in the inset of Fig.~\ref{Fig004}. Although by taking into account more diagrams, the DMFT always gives a greater binding energy than the MA, for $t\gg\omega_0$ the DMFT, just as the MA, wrongly predicts a sudden change in the slope of the ground-state energy.

Within the Holstein model the sudden change in the ground-state properties for $t\gg\omega_0$ occurs if the dimension of the system is greater than one. Namely, for $D>1$ the large adiabatic polarons are unstable irrespective of the parameters.\cite{HolEmin} Specifically, for $D>1$ and $t/\omega_0\gg1$, the weakly dressed electron (local self-energy) crosses\cite{Eagles} directly into a heavy nearly-small polaron (nearly local self-energy). Obviously, this kind of behavior which does not involve long-range adiabatic correlations is more likely to be correctly reproduced by the DMFT. Particularly in the $D\rightarrow\infty$ limit where the exact self-energy becomes local.\cite{Ciuchi}

For models with short-range interactions as for the Holstein model discussed here, strong adiabatic correlations develop for significant electron-phonon couplings, for which the electron spectral weight at low frequencies is strongly suppressed. Therefore, although the low-frequency dynamics of the charge carriers might be governed by the adiabatic correlations, one may find that these correlations are difficult to observe directly in experiments that measure the spectral function of the electron,\cite{Mishchenko} e.g., in photoemission or tunneling measurements. In such circumstances, the investigations that reveal the phonon properties,\cite{Bozin} which are strongly affected by polaron adiabatic correlations,\cite{Barisic3,Loos} might provide better insights in strongly coupled electron-phonon systems.

In conclusion, for the Holstein model the DMFT sums an infinite series of local diagrams characterized by the absence of vertex corrections involving different lattice sites. By analyzing the single-electron problem it is shown that the vertex corrections neglected by the DMFT are important in low dimensions for the description of the adiabatic polaronic correlations spreading over multiple lattice sites.


\begin{thebibliography}{99}

\bibitem{Pruschke} For a review, see T. Pruschke, M. Jarrell, and J. Freericks,
	Adv. Phys. {\bf 44}, 187 (1995);
	A. Georges, G. Kotliar, W. Krauth, and M. Rozenberg,
	Rev. Mod. Phys. 68, 13 (1996), and references therein.

\bibitem{Koller}
	W. Koller, D. Meyer, and A. C. Hewson,
	Phys. Rev. B {\bf 70}, 155103 (2004); 
	M. S. Laad, L. Craco, and E. M\" uller-Hartmann,
	{\it ibid.} {\bf 73}, 195120 (2006); 
	P. Werner and A. J. Millis,
	{\it ibid.} {\bf 75}, 085108 (2007).

\bibitem{Metzner} W. Metzner and D. Vollhardt,
	Phys. Rev. Lett. {\bf 62}, 324 (1989).

\bibitem{Metzner2} W. Metzner.
	Phys. Rev. B {\bf 43}, 8549 (1991).

\bibitem{Ciuchi} S. Ciuchi, F. de Pasquale, S. Fratini, and D. Feinberg,
	Phys. Rev. B {\bf 56}, 4494 (1997).

\bibitem{Paci} P. Paci, M. Capone, E. Cappelluti, S. Ciuchi, C. Grimaldi, and L. Pietronero,
	Phys. Rev. Lett. {\bf 94}, 036406 (2005); 
	W. Koller, A. C. Hewson, and D. M. Edwards,
	{\it ibid.} {\bf 95}, 256401 (2005); 
	S. Fratini and S. Ciuchi,
	Phys. Rev. B {\bf 74}, 075101 (2006). 
	
\bibitem{Filippis} G. Wellein and H. Fehske,
	Phys. Rev. B {\bf 56}, 4513 (1997);
	A. H. Romero, D. W. Brown, and K. Lindenberg,
	J. Chem. Phys. {\bf 109}, 6540 (1998);
	V. Cataudella, G. De Filippis, and G. Iadonisi,
	Phys. Rev. B {\bf 62}, 1496 (2000);
	Li-Chung Ku, S. A. Trugman, and J. Bon\v ca,
	{\it ibid.} {\bf 65}, 174306 (2002);
	P. E. Spencer, J. H. Samson, P. E. Kornilovitch, and A. S. Alexandrov,
	{\it ibid.} {\bf 71}, 184310 (2005).
	
\bibitem{Barisic5} O. S. Bari\v si\' c,
	Europhys. Lett. {\bf 77}, 57004 (2007).

\bibitem{Holstein} T. Holstein,
	Ann. Phys. (N.Y.) {\bf 8}, 325 (1959).

\bibitem{BBarisic} O. S. Bari\v si\' c and S. Bari\v si\' c, 
	Eur. Phys. J. B {\bf 54}, 1 (2006).

\bibitem{Barisic3} O. S. Bari\v si\' c,
	Phys. Rev. B {\bf 73}, 214304 (2006).

\bibitem{Barisic6} O. S. Bari\v si\' c, Phys. Rev. Lett. {\bf 98}, 209701 (2007).

\bibitem{Berciu} M. Berciu,
	Phys. Rev. Lett. {\bf 97}, 036402 (2006).

\bibitem{Goodvin} G. L. Goodvin, M. Berciu, and G. A. Sawatzky,
	Phys. Rev. B {\bf 74}, 245104 (2006).

\bibitem{Jarrell} M. Jarrell,
	Phys. Rev. Lett. {\bf 69}, 168 (1992).

\bibitem{Rashba} E. I. Rashba,
	Opt. Spectrosk. {\bf 2}, 664 (1957).

\bibitem{Mahan} G. D. Mahan,
	{\it Many-Particle Physics} (Plenum Press, New York, USA, 1990).

\bibitem{Emin} D. Emin,
	Phys. Rev. B {\bf 48}, 13691 (1993).

\bibitem{Berciu2} M. Berciu,
	Phys. Rev. Lett. {\bf 98}, 209702 (2007).

\bibitem{Maier} T. Maier, M. Jarrell, T. Pruschke, and M. H. Hettler,
	Rev. Mod. Phys. {\bf 77}, 1027 (2005).


\bibitem{HolEmin} D. Emin and T. Holstein,
	Phys. Rev. Lett. {\bf 36}, 323 (1976);
	G. Kalosakas, S. Aubry, and G. P. Tsironis,
	Phys. Rev. B {\bf 58}, 3094 (1998).

\bibitem{Eagles} D. M. Eagles,
	Phys. Rev. {\bf 145}, 645 (1966). 

\bibitem{Mishchenko} A. S. Mishchenko and N. Nagaosa,
	Phys. Rev. Lett. {\bf 93}, 036402 (2004);
	K. M. Shen, F. Ronning, W. Meevasana, D. H. Lu, N. J. C. Ingle, F. 	Baumberger, W. S. Lee, L. L. Miller, Y. Kohsaka, M. Azuma, M. Takano,
	H. Takagi, and Z. X. Shen,
	Phys. Rev. B {\bf 75}, 075115 (2007).

\bibitem{Bozin} N. Mannella, A. Rosenhahn, C. H. Booth, S. Marchesini, B. S. 	Mun, S.-H. Yang, K. Ibrahim, Y. Tomioka, and C. S. Fadley,
	Phys. Rev. Lett. {\bf 92}, 166401 (2004);
	E. S. Bo\v zin, M. Schmidt, A. J. DeConinck, G. Paglia, J. F. Mitchell,
	T. Chatterji, P. G. Radaelli, Th. Proffen, and S. J. L. Billinge,
	Phys. Rev. Lett. {\bf 98}, 137203 (2007).

\bibitem{Loos} J. Loos, M. Hohenadler, A. Alvermann, and H. Fehske,
	J. Phys.: Condens. Matter. {\bf 18}, 7299 (2006).

\end{thebibliography}
\end{document}